# Wafer-scale conformal metasurface optics

**Louis Martin-Monier[1], Sehui Chang[2,3], Johannes Froech[4], Zhaoyi Li[1], Luigi Ranno[1], Khoi Phuong Dao[1], Akira Ueno[1], Jia Xu Brian Sia[1], Hanyu Zheng[1,5], Padraic Burns[5], Tian Gu[1,5*], Young Min Song[2,3*], Arka Majumdar[4], and Juejun Hu[1,5]**

*[1]Department of Materials Science & Engineering, Massachusetts Institute of Technology, Cambridge, Massachusetts, USA*
*[2]School of Electrical Engineering, Korea Advanced Institute of Science and Technology, Daejeon, Republic of Korea*
*[3]GIST InnoCORE AI-Nano Convergence Institute for Early Detection of Neurodegenerative Diseases, Gwangju Institute of Science and Technology, Gwangju, Republic of Korea*
*[4]Department of Electrical and Computer Engineering, University of Washington, Seattle, Washington, USA*
*[5]2Pi Inc., Cambridge, Massachusetts, USA*

*gutian@mit.edu, ymsong@kaist.ac.kr

**Abstract**

Curved and conformal optics offer significant advantages by unlocking additional geometric degrees of freedom for optical design. These capabilities enable enhanced optical performance and are essential for meeting non-optical constraints, such as those imposed by ergonomics, aerodynamics, or wearability. However, existing fabrication techniques such as direct electron or laser beam writing on curved substrates, and soft-stamp-based transfer or nanoimprint lithography suffer from limitations in scalability, yield, geometry control, and alignment accuracy. Here, we present a scalable fabrication strategy for curved and conformal metasurface optics leveraging thermoforming, an industry-standard, high-throughput manufacturing process widely used for shaping thermoplastics. Our approach uniquely enables wafer-scale production of highly curved metasurface optics, achieving sub-millimeter radii of curvature and micron-level alignment precision. To guide the design and fabrication process, we developed a thermorheological model that accurately predicts and compensates for the large strains induced during thermoforming. This allows for precise control of metasurface geometry and preservation of optical function, yielding devices with diffraction-limited performance. As a demonstration, we implemented an artificial compound eye comprising freeform micro-metalens arrays. Compared to traditional micro-optical counterparts, the device exhibits an expanded field of view, reduced aberrations, and improved uniformity, highlighting the potential of thermoformed metasurfaces for next-generation optical systems.

## Main

Flat optics exemplified by metasurfaces and diffractive optical elements (DOEs) offer major size, weight, and cost advantages over traditional bulk optics and are seeing growing commercial deployment in recent years. Taking advantage of their ultrathin form factor, the "flat" optics can also assume non-planar forms, either as standalone elements or as a conformal layer integrated on curved substrates[1–3]. The added geometric degrees of freedom have been proven to significantly enhance their optical performance and functionality compared to flat counterparts[4–7]. For example, conformal integration of metasurfaces with freeform reflective[8] or refractive[9,10] optics can improve aberration correction, affording enhanced imaging quality with the same or even reduced number of optical elements. Moreover, separating the physical shape of meta-optical components from their optical function enables independent optimization for factors such as ergonomics[11], aerodynamics[12], and wearability[13].

Current fabrication methods for conformal metasurfaces and DOEs suffer from significant limitations in scalability, geometric control, and alignment accuracy. While laser[14] or electron beam[8] direct-write techniques have been adapted for patterning on curved substrates, their inherently low throughput renders them unsuitable for large-scale manufacturing. Alternative strategies employ elastomeric stamps capable of conforming to non-planar surfaces for direct nanoimprint lithography[15–17]. These direct patterning approaches are constrained to gently curved surfaces, with radii of curvature generally exceeding several centimeters. Another common approach involves fabricating metasurfaces on flat, rigid substrates followed by transfer to elastomeric carriers for conformal integration[18–21]. Precise alignment, which is crucial to assembly of optical systems involving these non-planar elements, is however challenging as elastomers are prone to mechanical deformation. Although non-elastomeric polymer substrates have also been used[22,23], their limited stretchability restricts the achievable curvature to one-dimensional bending.

To circumvent the limitations of the existing methods, we turn to thermoforming, an industry-proven technique renowned for its scalability and reliability in high-volume production of thermoplastic components. In thermoforming, a rigid thermoplastic sheet initially stiff and dimensionally stable at room temperature is heated to a pliable state where its viscosity falls within the range of $10^4$ to $10^6$ Pa·s. At this viscoelastic condition, the sheet can be shaped into a desired form using a mold through vacuum, pressure, or mechanical forming. Upon cooling, the material regains its original rigidity, permanently retaining the newly formed geometry. Inspired by the thermoforming process, we developed a curved and conformal metasurface fabrication strategy that is fully compatible with high-throughput wafer-foundry processes, including deep-ultraviolet (DUV) lithography and nanoimprint techniques. Our thermoforming-based method enables large-area shaping of conformal optics with high dimensional fidelity, even for geometries featuring tight, millimeter-scale curvature. The inherent mechanical rigidity and dimensional stability of the thermoplastic sheet further support precise wafer- or panel-level alignment, ensuring accurate pattern registration throughout the optics assembly process. Together, these advantages (Table 1) position our approach as a scalable and industrially viable solution next-generation conformal photonic systems.

### Thermoforming-based conformal metasurface fabrication

The fabrication process is schematically depicted in Fig. 1a. The process begins with the deposition and lithographic patterning of a high-refractive-index layer into metasurface structures on a flat wafer. In this study, hydrogenated amorphous silicon (a-Si) is used as the meta-atom material, though the overall approach is readily adaptable to other metasurface platforms. After a brief oxygen plasma treatment to improve surface wettability, a solution of thermoplastic polymer (polyvinyl chloride (PVC) dissolved in cyclohexanone in our study) is drop-cast and spin-coated onto the patterned wafer. Subsequent thermal annealing forms a thin encapsulation layer, typically ~ 2 μm thick, conformally covering the a-Si metasurface. A thermoplastic PVC sheet of tailored thickness (typically several hundred microns) is then thermally bonded onto this encapsulated layer. To release the structure, the thermoplastic-embedded metasurface is separated from the substrate using a wet etchant that selectively attacks the a-Si/glass interface. The rapid lateral etching enables clean delamination while minimizing chemical damage to the a-Si nanostructures.

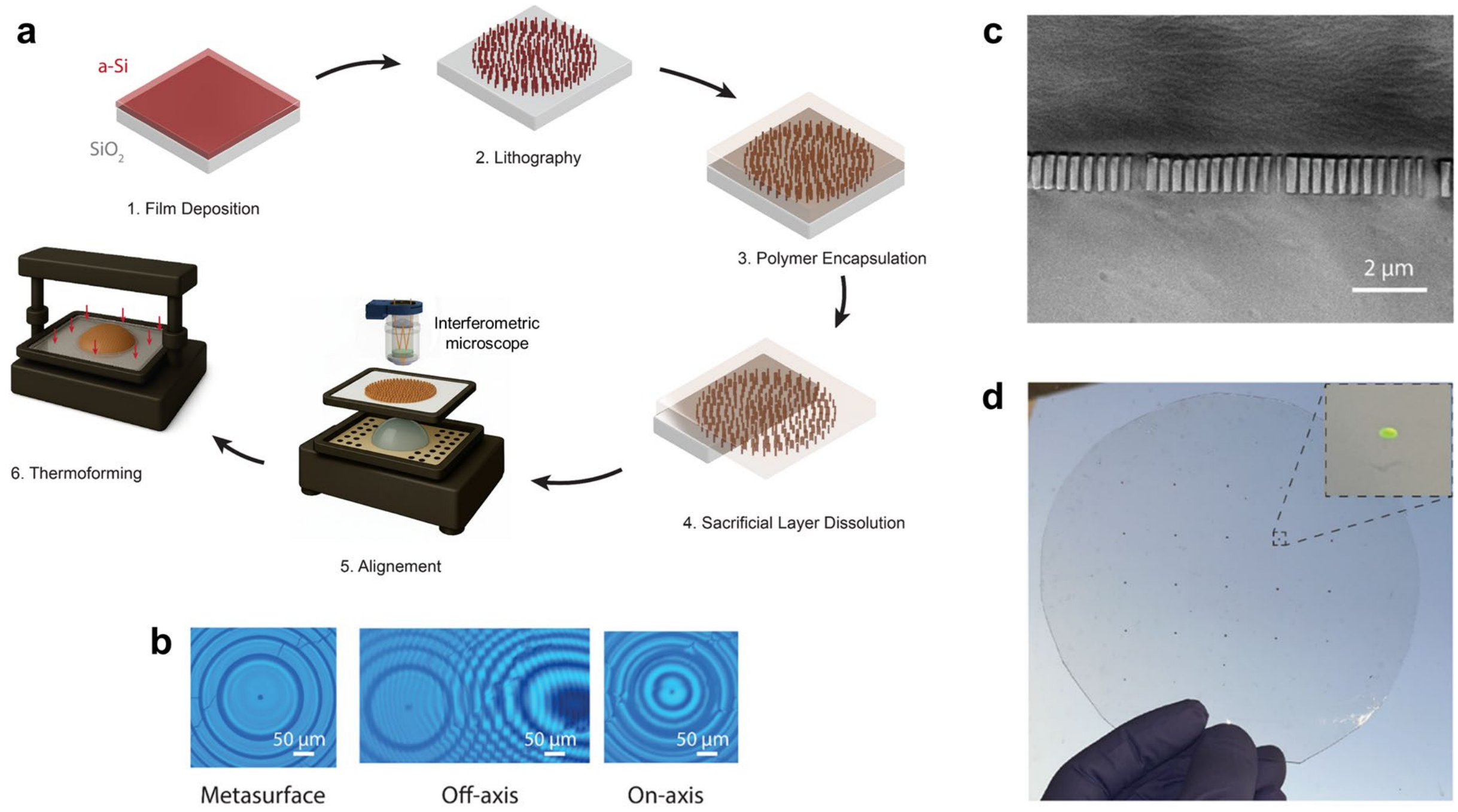

**Fig. 1. Conformal metasurface fabrication and assembly. (a) Schematic illustration of the fabrication process flow: (1) deposition of metasurface material (here a-Si) onto a handler substrate; (2) patterning of the metasurface using electron beam lithography and plasma etching; (3) encapsulation of the metasurface via spin-coating of a dissolved polymer, followed by thermal bonding to a thicker polymer support sheet of the same material; (4) release of the encapsulated structure by lateral wet etching; (5) alignment of the planar polymer-embedded metasurface over a curved mold using interferometric microscopy; and (6) thermoforming of the metasurface to conform to the mold surface. During alignment, a 480 nm monochromatic light source is used to generate Newton's rings, which serve as fiducial markers. The center of the metasurface phase profile, delineated by visible Fresnel zones, is aligned with the interference fringes to ensure concentric registration. (b) Interferometric microscope images showing (left) the metasurface suspended above the mold with no contact, (middle) misaligned contact between the metasurface and mold, and (right) optimal alignment indicated by concentric interference fringes overlapping with the Fresnel zones. (c) Cross-sectional SEM image of a curved metasurface after thermoforming, confirming structural integrity after the process. (d) Photograph of a 6-inch substrate with an embedded metasurface array fabricated using the proposed process, demonstrating large-area scalability of the process. Inset shows an individual curved metasurface.**

The released structure, now in the form of a rigid thermoplastic substrate, is subsequently shaped via the thermoforming process. Precise alignment between the metasurface-embedded substrate and the mold is essential for achieving high-quality optical performance. In this study, we developed an alignment protocol based on interferometric microscopy to ensure micron-level placement accuracy. Importantly, because the form factor and mechanical rigidity of the thermoplastic substrate closely match those of conventional glass wafers, standard wafer- or panel-level alignment techniques commonly used in semiconductor manufacturing can also be readily applied. The aligned substrate-mold assembly is then transferred to a commercial benchtop thermoforming machine, where the thermoplastic substrate is heated to a pliable state and vacuum is applied through the perforated base to conform it onto the mold surface. After cooling, the molded metasurface can either be demolded to serve as a freestanding optical element or remain permanently bonded to the rigid mold—such as a refractive lens—to create hybrid refractive-metasurface optics. Details of the fabrication process are furnished in Methods.

We demonstrate the scalability of our thermoforming process through a wafer-scale fabrication test. A sparse 5 × 5 array of metasurfaces, each occupying a 2 mm × 2 mm area, was patterned via electron beam lithography on a 6-inch glass wafer. The metasurface structures were subsequently transferred to a 6-inch PVC substrate and thermoformed using the previously described protocol. Off-the-shelf plano-convex lenses mounted on a base plate served as the thermoforming mold. Fig. 1b shows interferometric microscope images acquired during the alignment process. The right panel displays concentric Newton's rings centered on the metasurface, indicating optimal alignment with the mold, while the center panel shows misalignment. Using this method, we achieve alignment accuracy within ± 10 μm across the conformal optics, which, according to our simulations, introduces negligible degradation to the optical performance of the resulting curved metasurfaces. Importantly, this limit is imposed not by the alignment protocol itself but by non-uniform heating and post-forming thermal contraction; both effects could be further mitigated by employing thermoforming systems with enclosed, multi-zone heating to improve temperature uniformity. Fig. 1c presents a cross-sectional scanning electron microscopy (SEM) image of the molded metasurface, confirming that the meta-atoms retain their structural integrity after thermoforming. The apparent tapering of the meta-atom profiles is an artifact resulting from the focused ion beam (FIB) cross-sectioning plane being slightly tilted relative to the local surface normal of the curved polymer substrate. Fig. 1d shows a photograph of the fully processed PVC substrate containing a 5 × 5 array of thermoformed metasurfaces, highlighting the scalability of the process. Although electron beam lithography was employed in this proof-of-concept demonstration, the results clearly establish the compatibility of our thermoforming approach with wafer-scale metasurface fabrication workflows, underscoring its potential for large-area production of curved and conformal optical elements.

## A predictive thermorheological model for thermoforming metasurfaces

During the thermoforming process, the metasurface-embedded thermoplastic substrate undergoes substantial viscoelastic deformation. This deformation influences the optical performance of the resulting metasurfaces through two primary mechanisms: changes in the local substrate thickness and variation in the spacing between individual meta-atoms. Quantitative understanding and control of these effects are essential for achieving high-fidelity, optically functional curved and conformal metasurfaces. To this end, we developed a predictive thermorheological model that accurately captures the material response during thermoforming, enabling compensation for deformation-induced optical deviations.

We begin by modeling the thermoforming-induced deformation of a plain thermoplastic sheet—without embedded metastructures—using the Finite Element Method (FEM). The simulation incorporates experimentally measured, temperature-dependent rheological properties of the thermoplastic material, including its shear and loss moduli, as input parameters (Supplementary Note 1). Further simulation details are provided in the Methods section and Supplementary Note 2. Fig. 2a plots the predicted surface strain distribution along the radial direction at the top surface of a 200 μm-thick plastic sheet thermoformed over a rotationally symmetric mold with a radius of curvature $R$ = 3.2 mm at 100 °C. To validate the model, we experimentally measured local deformation by tracking the displacement of a sparse rectangular array of silicon islands embedded in the polymer film. The measured strain closely matches the FEM predictions, confirming the accuracy of our modeling framework. Furthermore, surface strain profiles measured along three orthogonal radial directions (Fig. 2b) reveal excellent symmetry, indicating that the thermoforming process induces uniform, isotropic deformation. This observation validates the model's assumption of axisymmetry and supports the absence of process-induced anisotropy in the resulting structures.

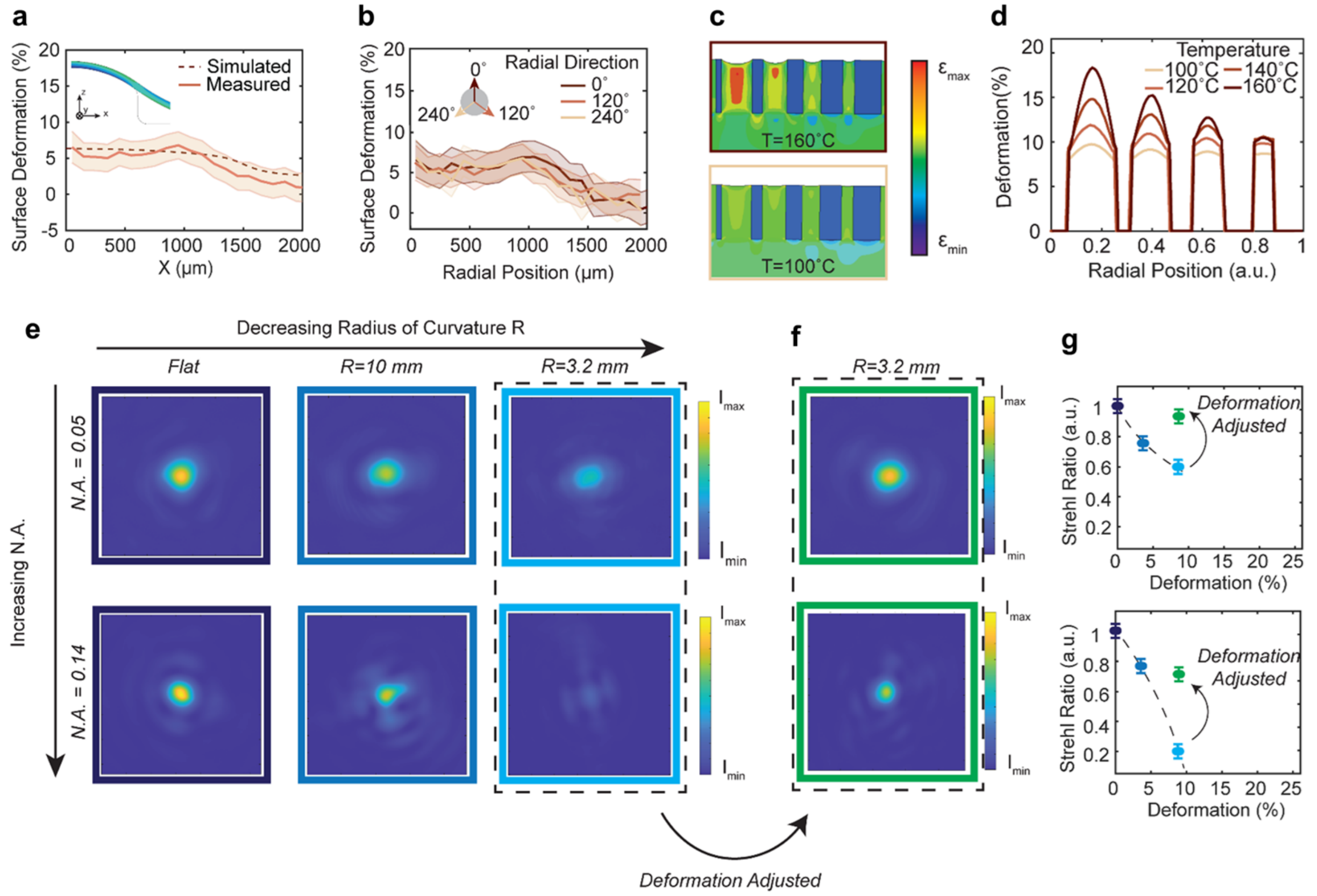

**Fig. 2. Optomechanical behavior of thermoformed curved metasurfaces. (a) Simulated (dashed line) and measured (solid line) surface deformation of a 200 μm thick polymer sheet thermoformed over a spherical cap of radius of curvature 3.2 mm at 100 °C, using experimentally extracted rheological parameters. Inset: cross-section of the deformed film profile. (b) Surface deformation profiles measured along three radial directions, confirming the axisymmetric nature of the deformation process. Shaded regions represent experimental error. (c) Finite element simulation of radial strain distribution across a single Fresnel zone composed of five meta-atom pillars, at two forming temperatures. At 100 °C (bottom), strain remains symmetric around each meta-atom; at 160 °C (top), strain asymmetry emerges due to enhanced flow and redistribution. (d) Simulated radial deformation profiles under various temperatures, showing that lower thermoforming temperatures result in more uniform strain distributions. (e) Measured PSFs for metalenses of increasing numerical aperture and decreasing radius of curvature without deformation compensation. (f) Measured PSFs of curved metalenses incorporating deformation-compensated design. (g) Simulated (dotted lines) and measured (symbols) Strehl ratios, plotted as a function of deformation. Our design accounting for the deformation effect recovers near-diffraction-limited performance.**

We further extend our model by incorporating embedded a-Si meta-atoms (assumed to be mechanically rigid) into the thermoplastic substrate to comprehensively characterize thermoforming-induced deformation. To investigate their influence on local strain behavior, we vary the diameters and spacings of the meta-atoms. Fig. 2c and Fig. 2d show simulated strain profiles for a representative metasurface segment corresponding to a single Fresnel zone at different thermoforming temperatures. The results indicate that spatial variations in the volume fill factor of meta-atoms give rise to non-uniform strain distributions, which in turn lead to disproportionate changes in inter-atom spacing across the metasurface. Such strain non-uniformity becomes more pronounced at elevated temperatures, consistent with the decrease in polymer viscosity and the associated increase in local stress redistribution.

We next present our design strategy, which leverages model-predicted strain distributions to adapt the metasurface layout for optimal optical performance under thermoforming-induced deformation. Thermoforming affects metasurface behavior through four principal mechanisms: (1) lateral displacement of meta-atoms, which perturbs the intended optical phase profile; (2) variation in substrate thickness, which modifies local optical path lengths; (3) changes in inter-atom spacing, which influence near-field coupling and alter the optical response of individual meta-atoms; and (4) modification of the local angle of incidence due to surface curvature. To illustrate the impact of uncorrected deformation, we modeled two metalenses—each embedded in a PVC substrate and initially designed for a flat geometry—operating at a wavelength of 940 nm with numerical apertures (NAs) of 0.05 and 0.14, respectively. The design parameters of the meta-atoms and their responses to both in-plane strain and bending-induced deformation are detailed in Supplementary Note 3. Fig. 2g shows that the simulated Strehl ratio of the uncorrected metalens decreases monotonically with increasing surface deformation (dotted line), consistent with experimental measurements (symbols). As expected, metalenses with higher NA exhibit greater susceptibility to performance degradation due to their higher phase gradients, which amplify the effects of deformation-induced phase errors. This degradation is further visualized in Fig. 2e, which shows the measured point spread functions (PSFs) for the two NA designs across increasing curvature conditions. The results highlight the substantial loss of focusing quality at small radii of curvature, particularly for the higher NA design. These findings underscore the importance of incorporating deformation-aware corrections into the metasurface design to preserve optical performance under curved configurations.

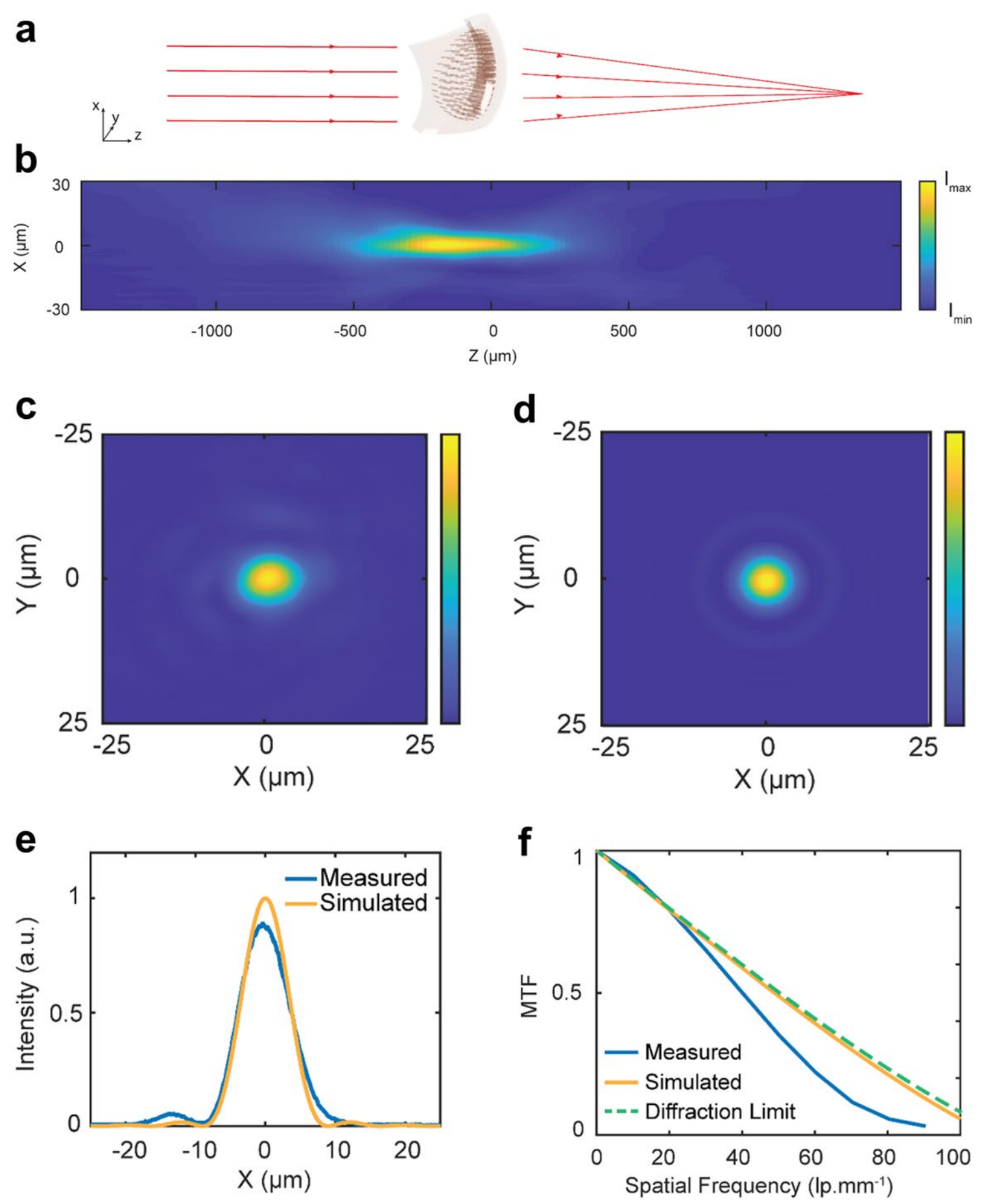

**Fig. 3. Experimental characterization of a freestanding curved metalens. (a) Schematic illustrating the operation of the curved metalens under collimated monochromatic illumination. (b) Measured optical intensity distribution along the optical (*z*) axis, captured in the *x*-*z* plane. (c-d) PSF intensity maps at the focal plane (f = 7 mm): (c) experimental measurement and (d) simulated reference, showing close agreement. (e) PSF line profile along the *x*-direction at the focal plane, comparing experimental data (blue) and simulation (orange). (f) MTF as a function of spatial frequency: (green dashed line) theoretical diffraction limit; (orange line) simulated MTF for the lens; (blue line) experimentally retrieved MTF obtained via inverse Fourier transform of the measured PSF. The results confirm near-diffraction-limited performance of the curved metalens.**

To mitigate the performance degradation induced by thermoforming, we implemented a multi-faceted compensation strategy informed by FEM model predictions. Specifically, we adjusted the initial positions of the meta-atoms to offset their displacement during forming and modified the optical phase profile to account for changes in substrate thickness. In parallel, we developed meta-atom designs that are robust to variations in both incident angle and inter-atom spacing, thereby minimizing sensitivity to deformation-induced perturbations. Fig. 2f presents the PSFs of two metalenses incorporating this corrected design. The restored near-diffraction-limited Strehl ratios confirm the effectiveness of our compensation strategy in preserving optical performance despite the large mechanical

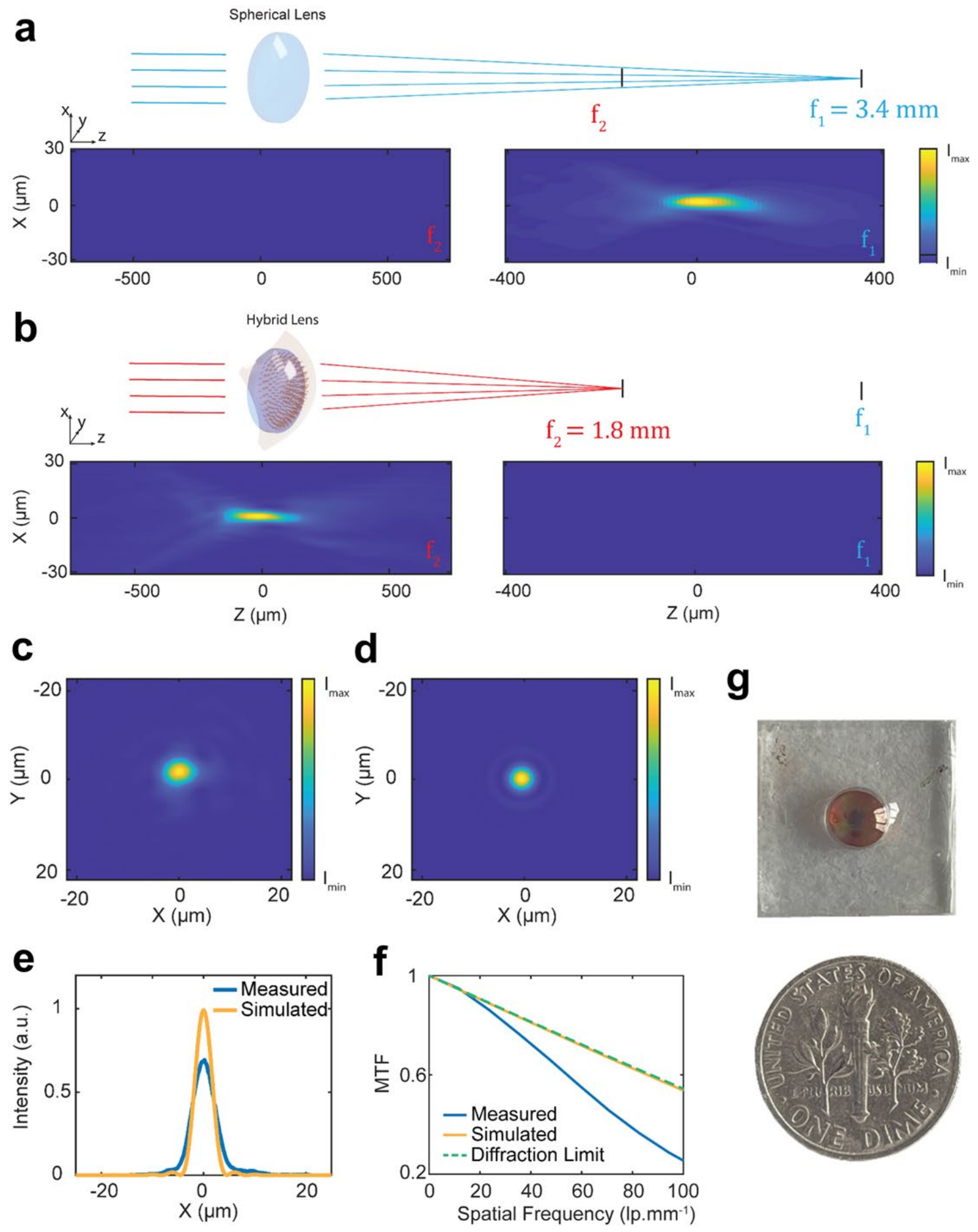

**Fig. 4. Experimental demonstration of a hybrid refractive-metasurface optic. (a-b) Schematics and measured intensity distributions along the optical axis under collimated monochromatic illumination for (a) a plano-convex spherical lens (focal length $f_1$ = 3.4 mm, NA = 0.10) and (b) a hybrid optic consisting of the same spherical lens conformally integrated with a thermoformed metalens, reducing the effective focal length to $f_2$ = 1.8 mm, (NA = 0.18). (c-d) Cross-sectional intensity distributions of the PSF at the focal plane of the hybrid optic: (c) experimental measurement and (d) simulation, showing close agreement. (e) PSF line scan across the focal spot, comparing experiment (blue) and simulation (orange). (f) MTF of the hybrid optic: (green dashed line) diffraction limit; (orange line) simulated performance; (blue line) experimental MTF obtained via Fourier transform of the measured PSF. (g) Photograph of the hybrid optic shown next to a U.S. dime for scale.**

deformations introduced during thermoforming.

## Freestanding curved metalens characterization

We evaluated the optical performance of a freestanding curved metalens with a numerical aperture (NA) of 0.14 and a radius of curvature $R = 3.2$ mm, designed to operate at a wavelength of 940 nm and fabricated using the thermoforming method. The deformation compensation strategy described in the previous section was applied during the design process. Fig. 3a shows a schematic of the device under collimated illumination. To assess focusing performance, we experimentally characterized the 3D PSF of the metalens. Fig. 3b and Fig. 3c present the measured intensity distributions in the *x-z* and *x-y* planes, respectively, alongside simulation of an ideal curved lens without fabrication imperfections (Fig. 3d). The metalens demonstrates diffraction-limited performance, with a Strehl ratio of 0.89 (Fig. 3e) and an MTF closely matching that of the ideal aberration-free design (Fig. 3f).

## Conformal refractive-metasurface hybrid lens characterization

As a next demonstration, we evaluated a conformal refractive-metasurface hybrid lens, fabricated using an off-the-shelf plano-convex lens as the mold. Through thermoforming, the metasurface layer was conformally bonded onto the curved surface of the refractive lens, forming an integrated hybrid optical system. Such hybrid optics offer unique performance advantages by combining the precise monochromatic aberration correction enabled by metasurfaces with the large group delay inherent to refractive lenses—critical for broadband achromatic operation.

As a proof-of-concept, we fabricated a hybrid lens shown in Fig. 4g, in which the metasurface is designed to enhance the refractive power of the underlying lens. The bare refractive lens depicted in Fig. 4a has a focal length of $f_1 = 3.4$ mm, which is reduced to $f_2 = 1.8$ mm after metasurface integration. Experimentally measured intensity distributions around both focal planes are presented in the lower panels of Fig. 4a and Fig. 4b. The hybrid lens exhibits a peak intensity ratio of $I_1 / I_2 = 80$ at the two focal planes $f_1$ and $f_2$, demonstrating exceptionally high contrast between the desired (first-order) and residual (zeroth-order) diffraction of the conformal metasurface. This result underscores the effectiveness of our fabrication and deformation-compensation strategies. We further characterized the hybrid lens's focusing performance at the $f_2$ focal plane. The measured PSF (Fig. 4c) shows good agreement with simulations (Fig. 4d), yielding a Strehl ratio of 0.72 according to Fig. 4e. The results indicate that the conformal hybrid lens operates near the diffraction limit.

## Curved metalens compound eye imaging

Insect compound eyes have long inspired ultra-compact artificial vision systems due to their distinctive micro-optic array architecture (Fig. 5a). Compared to planar counterparts, curved compound eyes provide a wider field-of-view with reduced aberrations, making them particularly advantageous for compact wide-angle imaging applications[24]. Conventional microlens-based curved artificial compound eyes fabricated using direct laser writing or pneumatic deformation often suffer from non-uniform lens deformation and limited scalability when adapted to curved geometries, ultimately degrading imaging performance[25]. To overcome these limitations, we fabricated a curved metalens compound eye by thermoforming a flat metalens array (Fig. 5b-c). This approach enables precise control of curvature across the entire array, ensuring uniform focusing behavior and consistent image quality from the central to peripheral regions.

We conducted imaging demonstrations with the curved metalens compound eye to evaluate its ability to form high-quality compound-eye images across the entire array. The system successfully generated well-focused images of a checkerboard target at working distances ranging from 2 cm to 10 cm, confirming its large depth-of-field (Fig. 5d). To further test angular performance, we established an optical setup for object imaging under varying incident angles, as detailed in the Methods section. The curved metalens compound eye produced uniform subaperture images of the letter "E" at incident angles of 0°, 15°, and 30° (Fig. 5e-g). Notably, the contrast profiles shown in Fig. 5h remained consistent across all angles, verifying that conformal fabrication preserved the optical performance of metalenses in both the central and peripheral regions.

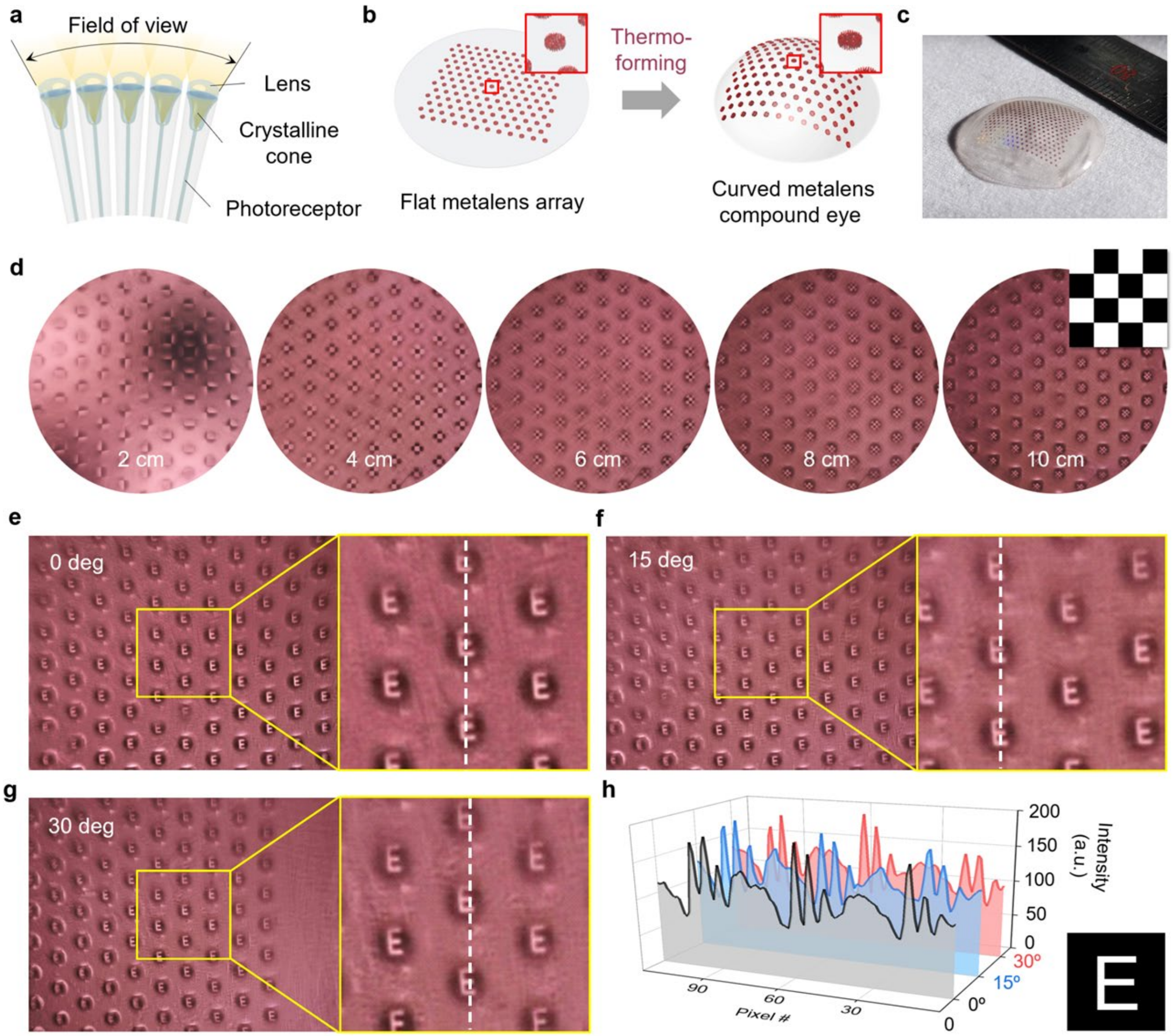

**Fig. 5. Imaging demonstration of a conformal metalens compound eye. (a) Schematic of an insect compound eye, consisting of multiple microlenses coupled to crystalline cones and photoreceptors, enabling a wide field-of-view. (b) Schematic of the thermoforming process, transforming a flat metalens array into a curved metalens compound eye. (c) Photograph of the fabricated curved metalens compound eye, shown next to a ruler for scale. (d) Imaging results of the curved metalens compound eye for a checkerboard target at varying working distances. (e-g) Imaging results of the letter "E" under different incident angles: (e) 0°, (f) 15°, and (g) 30°. Insets show magnified views of the captured images, with dotted lines marking the locations used for intensity profiling. (h) Line intensity profiles extracted along the dotted lines in (e-g), demonstrating consistent image contrast across both central and peripheral regions of the array, confirming uniform optical performance of the compound eye.**

## Conclusions

We have developed and experimentally validated a thermoforming-based strategy for the fabrication of curved and conformal metasurfaces. Leveraging an industry-standard process, our approach achieves wafer-scale scalability while maintaining micron-level alignment accuracy and sub-millimeter radii of curvature. Through systematic thermorheological modeling, we established a predictive framework to compensate for deformation-induced optical errors, enabling devices with diffraction-limited performance.

We demonstrated the versatility of this platform by realizing freestanding curved metalenses, conformal refractive-metasurface hybrid optics, and a bioinspired compound eye. In each case, our results highlight the ability to preserve structural integrity, achieve high focusing fidelity, and deliver uniform imaging performance across the entire aperture. Importantly, the use of rigid thermoplastic sheets ensures compatibility with standard wafer-level lithography, including DUV and nanoimprint techniques, thus providing a direct pathway toward large-area, high-throughput manufacturing.

Compared to existing approaches based on direct writing, soft stamping, or elastomer transfer, thermoforming uniquely combines precision, robustness, and scalability. By decoupling optical design from geometric constraints, our method opens opportunities for integrating conformal metasurfaces into compact, wide-angle, and multifunctional optical systems. The demonstrated wafer-scale compatibility establishes thermoforming as a practical and industrially viable solution for next-generation conformal photonics, with applications spanning imaging, sensing, wearable displays, and bioinspired vision systems.

**Table 1. Representative experimental demonstrations of curved and conformal metasurface optics fabrication methods. N/A indicates that the corresponding information was not provided in the referenced publication.**

| | **Fabrication method** | **Sample area/size** | **Freestanding?** | **2D conformal integration?** | **Smallest radius of curvature** | **Alignment accuracy** |
|---|---|---|---|---|---|---|
| *Adv. Photonics. Res.* **5**, 2300241 (2024)[14] | Direct laser writing | ~12.7 mm diameter lens | No | Yes | 56.1 mm | N/A |
| *Sci. Adv.* **7**, eabe5112 (2021)[8] | Electron-beam lithography | 3 $mm^2$ patterned aperture on 6 mm diameter concave mirror | No | Yes | ~ 8 mm | Micron-level (alignment to fiducials) |
| *Science* **349**, 1310-1314 (2015)[26] | Electron-beam lithography on surface topology carved by focused ion beam | ~ 30 µm | No | Yes | Tens of microns | Sub-micron |
| *Nano. Res.* **11**, 2705 (2018)[15] | Nanoimprint | 35 mm diameter | No | Yes (gentle curves only) | 50 mm (spherical) | N/A |
| *Appl. Mater. Interfaces* 11, 26109-26115 (2019)[27] | Nanoimprint | ~ 3 cm diameter | Yes | Yes | Several centimeters | N/A |
| *PhotoniX* **4**, 18 (2023)[17] | Nanoimprint | ~ 5 cm diameter | Yes | Yes | ~ 10 cm | N/A |
| *Nat. Commun.* **7**, 11618 (2016)[18] | Transfer to elastomer substrates | A few $mm^2$ | No (flexible film requires support) | Yes | 4.13 mm (cylindrical) | Manual placement |
| *ACS Photonics* **5**, 1762-1766 (2018)[28] | Release of epoxy layer from flat substrate | Few-mm patch on curved object | No (film is flexible, but not in self-supporting shape) | No (bendable film, not stretchable) | 6 mm (cylindrical) | Manual placement |
| *Laser Photon Rev* **19**, 2401240 (2024)[11] | Release of epoxy layer from flat substrate | Few-mm patch on curved eyeglass | Yes | No (bendable film, not stretchable) | cm-scale | Manual placement |
| *Appl. Phys. Lett.* **123**, 231702 (2023)[29] | Release of epoxy layer from flat substrate | cm-scale films | Yes | No (bendable film, not stretchable) | 26 mm | N/A |
| *Adv. Sci.* **11**, 2407045 (2024)[30] | Transfer printing | cm-scale contact lens | No | Yes | ~ 9 mm | Manual placement |
| This work | Transfer to thermoplastic substrates followed by thermoforming | 6” diameter substrate (wafer-scale) | Yes | Yes | 3.2 mm | 10 µm |

## Methods

**Metasurface fabrication.** The fabrication process begins with a (100)-oriented silicon wafer (University Wafers, Inc.) as the substrate. A 180 nm-thick phosphate glass layer (Spin-on-Glass P-280HP, Desert Silicon) is spin-coated onto the wafer and annealed at 220 °C for 10 minutes. Subsequently, 100 nm of $SiO_2$ and 975 nm of a-Si are deposited via plasma-enhanced chemical vapor deposition (SAMCO, Inc.) at 350 °C and 270 °C, respectively. The metasurface pattern is defined using electron beam lithography (HS-50, STS-Elionix) operated at 50 kV beam voltage and 10 nA current. The pattern is written onto a 320 nm-thick ma-N 2403 negative-tone electron beam resist (Microresist Technology), spin-coated at 3000 rpm for 1 minute and soft-baked at 90 °C for 2 minutes. Prior to resist application, the substrate is treated with hexamethyldisilazane (HMDS) vapor to enhance adhesion. To suppress charging effects during exposure, a layer of water-soluble conductive polymer (ESpacer 300Z, Showa Denko America, Inc.) is applied over the resist. Post-exposure, the resist is developed in a tetramethylammonium hydroxide (TMAH)-based developer (AZ 726 MIF, Microchemicals). Pattern transfer into the a-Si layer is achieved via reactive ion etching (RIE; Pegasus, SPTS Technologies) using alternating cycles of $SF_6$ and $C_4F_8$. After etching, the remaining resist is removed by oxygen plasma ashing. For encapsulation, a PVC solution is prepared by dissolving 10 g of polyvinylchloride (PVC) in 100 mL of cyclohexanone. This solution is spin-coated onto the metasurface at 1500 rpm for 40 seconds with a ramp rate of 100 rpm/s, followed by annealing at 130 °C for 5 minutes. The encapsulation is repeated using a higher concentration solution (20 g PVC in 100 mL cyclohexanone), resulting in a total encapsulation layer thickness of approximately 10 µm. A PVC sheet (Sigma-Aldrich, Inc.) is then thermally bonded onto the encapsulated surface by annealing at 150 °C. To release the structure, the sample is immersed in a 1:1 mixture of 25% hydrofluoric acid (HF) and isopropyl alcohol (IPA), which selectively dissolves the underlying phosphate glass and $SiO_2$ layers. Once delaminated, the freestanding metasurface-embedded PVC sheet is rinsed with deionized water, blow-dried, and allowed to dry completely under ambient conditions.

**Thermoforming.** Prior to thermoforming, the thermoplastic substrate is mounted onto a concentric metal support ring (Xometry Inc.) to ensure mechanical stability and handling ease. This clamped assembly is then positioned above a rigid, curved mold that is bonded using UV-curable adhesive (OrmoComp, MicroResist GmbH) to a perforated metal base designed to enable vacuum suction during the forming process. The thermoplastic sheet and mold are initially brought into near contact without applying pressure. The entire setup is placed under an interferometric microscope equipped with a monochromatic blue light source ($\lambda = 480$ nm), a 50× interferometric magnification objective and a blue filter (center wavelength 488 nm, FWHM 10 nm, Thorlabs Inc.). The position of the mold base is finely adjusted relative to the support ring until concentric Newton's rings are centered over the metasurface region, indicating precise lateral alignment. Once alignment is achieved, the relative positions of the base and support ring are fixed using either mechanical clamps or UV-curable adhesive. The aligned assembly is then transferred to a commercial benchtop thermoforming machine (FormBox, Mayku Inc.). The PVC substrate is gradually heated to the target thermoforming temperature at a controlled rate of approximately 3 °C/min. After reaching the desired temperature, the system is held isothermally for 5 minutes to ensure thermal equilibrium. The block assembly is then lowered onto the mold, and a vacuum in the range of 5-10 kPa is applied for an additional 5 minutes to conform the softened PVC substrate to the mold surface. Following forming, the assembly is allowed to cool naturally to room temperature under ambient conditions.

**Thermorheological characterization and modeling.** To simulate microstructural deformation during thermoforming, we first experimentally measured the rheological properties of the PVC substrate. Dynamic mechanical analysis (ARES-G2, TA Instruments) was performed using 2° cone-plate geometry to obtain the storage and loss moduli of the PVC sheet. Angular frequency sweeps were conducted over the range of 0.3-100 rad/s following a 5-minute thermal stabilization period. The resulting viscoelastic parameters were fitted using the MCalibration software (PolymerFEM, Ansys Inc.) based on a linear elastic model with a fifth-order Prony series representation.

The deformation behavior was simulated using a dynamic implicit solver implemented in the Abaqus finite element modeling package (Simulia, Dassault Systèmes). Exploiting the axial symmetry of the system, a 2-D cross-sectional model of the thermoforming setup—including the PVC substrate, mold, support ring, and metal base—was constructed. The mold and metal base were modeled as rigid wire bodies fixed at reference points, while the PVC sheet was represented as a shell element, constrained at its contact point with the mold apex (initial alignment point) and at the inner edge of the concentric support ring. A vacuum load of 8 kPa was applied over a 5-second ramp, unless otherwise specified. Tangential contact interactions were defined with a standard friction coefficient of 0.3, while normal contact behavior was modeled as hard contact. A surface-to-surface discretization scheme was employed using a finite-sliding formulation, with contact stabilization enabled and a relative penetration tolerance set to 0.001. Deformation profiles were extracted along the radial direction and mapped against corresponding radial coordinates. For spatial resolution, a mesh size of 10 μm was used for the bulk substrate, while a refined mesh of 10 nm was applied in regions capturing microstructural deformation near the metasurface.

**Optical measurement.** Optical measurements were conducted using a 940 nm continuous-wave laser diode module (LQC940-90E, Newport Inc.) as the light source. The laser beam was first expanded using a beam expander (ZBE22, Thorlabs Inc.), then directed onto the conformal optics sample, which was mounted on a three-axis translational stage equipped with a two-angle tilt mount for precise orientation adjustment. The focused light from the sample was collected using a standard 50× microscope objective and projected onto a high-resolution near-infrared camera (Alvium 1800 U-501m NIR, Allied Vision Inc., 2592 × 1944 pixels). The camera and objective assembly were mounted on a separate three-axis translational stage. To obtain the three-dimensional optical intensity distribution, the $z$-position of the detection assembly was incrementally swept, capturing a sequence of $x$-$y$ sections at different depths to reconstruct the 3-D PSF.

The compound eye imaging experiments were conducted using a 780 nm collimated laser diode module (CPS780S, Thorlabs Inc.) with a plano-concave lens with a focal length of 150 mm (KPC031, Newport Corporation) to expand the light beam fully illuminating the object mask. A ground-glass diffuser (#47-953, Edmund Optics Inc.) was inserted between the lens and the object mask to ensure uniform illumination. The focused metalens images from the curved metalens compound eye were relayed via the 1:1 relay lens (MAP107575-A, Thorlabs Inc.) onto a CMOS image sensor (Sony IMX477R, 12.3MP, Sony Semiconductor Solutions Corporation). Imaging under different incident angles was carried out by rotating the object relative to the central curvature point of the curved metalens compound eye, ensuring that the rotation axis coincided with the array's geometric center.

## References

1. Zhou, Y. *et al.* Flexible Metasurfaces for Multifunctional Interfaces. *ACS Nano* **18**, 2685–2707 (2024).
2. Walia, S. *et al.* Flexible metasurfaces and metamaterials: A review of materials and fabrication processes at micro- and nano-scales. *Appl Phys Rev* **2**, 011303 (2015).
3. Geiger, S. *et al.* Flexible and Stretchable Photonics: The Next Stretch of Opportunities. *ACS Photonics* **7**, 2618–2635 (2020).
4. Burckel, D. B., Sweatt, W. C. & Reinke, C. M. Exploring curved metasurface optics. in *2024 IEEE Research and Applications of Photonics in Defense Conference, RAPID 2024 - Proceedings* (Institute of Electrical and Electronics Engineers Inc., 2024). doi:10.1109/RAPID60772.2024.10646914.
5. Aieta, F., Genevet, P., Kats, M. & Capasso, F. Aberrations of flat lenses and aplanatic metasurfaces. *Opt Express* **21**, 31530–31539 (2013).
6. Han, N., Huang, L. & Wang, Y. Illusion and cloaking using dielectric conformal metasurfaces. *Opt Express* **26**, 31625–31635 (2018).
7. Wu, K., Coquet, P., Wang, Q. J. & Genevet, P. Modelling of free-form conformal metasurfaces. *Nat Commun* **9**, 3494 (2018).
8. Nikolov, D. K. *et al.* Metaform optics: Bridging nanophotonics and freeform optics. *Sci Adv* **7**, eabe5112 (2021).
9. Yang, F. *et al.* Wide field-of-view metalens: a tutorial. *Advanced Photonics* **5**, 033001 (2023).
10. Cheng, J., Jafar-Zanjani, S. & Mosallaei, H. All-dielectric ultrathin conformal metasurfaces: lensing and cloaking applications at 532 nm wavelength. *Sci Rep* **6**, 38440 (2016).
11. Gan, Y. *et al.* See-Through Conformable Holographic Metasurface Patches for Augmented Reality. *Laser Photon Rev* **19**, 2401240 (2024).
12. Zhang, W., Zuo, B., Chen, S., Xiao, H. & Fan, Z. Design of fixed correctors used in conformal optical system based on diffractive optical elements. *Appl Opt* **52**, 461–466 (2013).
13. Wang, Y. *et al.* Wearable plasmonic-metasurface sensor for noninvasive and universal molecular fingerprint detection on biointerfaces. *Sci Adv* **7**, eabe4553 (2021).
14. Kang, M. *et al.* Photochemically Engineered Large-Area Arsenic Sulfide Micro-Gratings for Hybrid Diffractive–Refractive Infrared Platforms. *Adv Photonics Res* **5**, 2300241 (2024).
15. Bhingardive, V., Menahem, L. & Schvartzman, M. Soft thermal nanoimprint lithography using a nanocomposite mold. *Nano Res* **11**, 2705–2714 (2018).
16. Kim, G. *et al.* Metasurface-driven full-space structured light for three-dimensional imaging. *Nat Commun* **13**, 5920 (2022).
17. Choi, H. *et al.* Realization of high aspect ratio metalenses by facile nanoimprint lithography using water-soluble stamps. *PhotoniX* **4**, 18 (2023).
18. Kamali, S. M., Arbabi, A., Arbabi, E., Horie, Y. & Faraon, A. Decoupling optical function and geometrical form using conformal flexible dielectric metasurfaces. *Nat Commun* **7**, 11618 (2016).
19. Shi, Y. *et al.* Flexible Multilayer Metasurfaces: Design and Applications on Foldable and Conformable Substrates. *Adv Mater Technol* **10**, 2402130 (2025).
20. Kamali, S. M., Arbabi, E., Arbabi, A., Horie, Y. & Faraon, A. Highly tunable elastic dielectric metasurface lenses. *Laser Photon Rev* **10**, 1002–1008 (2016).

21. Ee, H.-S. & Agarwal, R. Tunable Metasurface and Flat Optical Zoom Lens on a Stretchable Substrate. *Nano Lett* **16**, 2818–2823 (2016).
22. Burch, J. & Di Falco, A. Holography Using Curved Metasurfaces. *Photonics* **6**, 8 (2019).
23. Zhang, X., Cai, H., Daqiqeh Rezaei, S., Rosenmann, D. & Lopez, D. A universal metasurface transfer technique for heterogeneous integration. *Nanophotonics* **12**, 1633–1642 (2023).
24. Lee, G. J., Choi, C., Kim, D. H. & Song, Y. M. Bioinspired Artificial Eyes: Optic Components, Digital Cameras, and Visual Prostheses. *Adv Funct Mater* **28**, 1705202 (2018).
25. Jiang, H. *et al.* Biomimetic Curved Artificial Compound Eyes: A Review. *Advanced Devices and Instrumentation* **5**, 0034 (2024).
26. Ni, X., Wong, Z. J., Mrejen, M., Wang, Y. & Zhang, X. An ultrathin invisibility skin cloak for visible light. *Science (1979)* **349**, 1310–1314 (2015).
27. Kim, K., Yoon, G., Baek, S., Rho, J. & Lee, H. Facile Nanocasting of Dielectric Metasurfaces with Sub-100 nm Resolution. *ACS Appl Mater Interfaces* **11**, 26109–26115 (2019).
28. Burch, J. & Di Falco, A. Surface Topology Specific Metasurface Holograms. *ACS Photonics* **5**, 1762–1766 (2018).
29. Biabanifard, M., Xiao, J. & Di Falco, A. Thin-film polymeric metasurfaces for visible wavelengths. *Appl Phys Lett* **123**, 231702 (2023).
30. Ko, J. *et al.* Metasurface-Embedded Contact Lenses for Holographic Light Projection. *Advanced Science* **11**, 2407045 (2024).
31. Shu, T., Pei, C., Wu, R., Li, H. & Liu, X. Design and fabricate freeform holographic optical elements on curved optical surfaces using holographic printing. *Opt Lett* **48**, 6537–6540 (2023).

**Acknowledgments**

This work was sponsored by the DARPA ENVision program. The views, opinions and/or findings expressed are those of the authors and should not be interpreted as representing the official views or policies of the Department of Defense or the U.S. Government.

**Author contributions**

L.M. performed finite element modelling, developed fabrication protocols, and characterized the metasurface optics. S.C. designed and characterized the compound eye micro-metalens array. J.F., L.R., K.P.D., A.U., and J.X.B.S. contributed to device fabrication. Z.L., H.Z., and P.B. assisted with metasurface optics testing. J.H. conceived the study. T.G., Y.M.S., A.M., and J.H. supervised and coordinated the research. J.H., L.M., S.C., and Y.M.S. drafted the manuscript. All authors contributed to technical discussions and writing the paper.

**Materials & Correspondence**

Correspondence and requests for materials should be addressed to Tian Gu and Young Min Song.

**Competing financial interests**

A patent based on the technology described herein has been filed and licensed to 2Pi Inc.